\documentclass[aps,prl,reprint,superscriptaddress,showpacs,amsmath,amssymb,longbibliography,lengthcheck]{revtex4-1}
\usepackage{graphicx}
\usepackage{mathrsfs}
\usepackage{txfonts}

\begin{document}

%Title of paper
 \title{Existence of exotic torus configuration in high-spin excited states of $^{40}$Ca} 

\author{T. Ichikawa}%
\affiliation{Yukawa Institute for Theoretical Physics, Kyoto University,
Kyoto 606-8502, Japan}
\author{J. A. Maruhn}
\affiliation{Institut fuer Theoretische Physik, 
Universitaet Frankfurt, D-60438 Frankfurt, Germany}
\author{N. Itagaki}
\affiliation{Yukawa Institute for Theoretical Physics, Kyoto University,
Kyoto 606-8502, Japan}
\author{K. Matsuyanagi}
\affiliation{Yukawa Institute for Theoretical Physics, Kyoto University,
Kyoto 606-8502, Japan}
\affiliation{RIKEN Nishina Center, Wako 351-0198, Japan}
\author{P.-G. Reinhard}
\affiliation{Institut f\"ur Theoretische Physik, Universit\"at Erlangen,
D-91058 Erlangen, Germany} 
\author{S. Ohkubo}
\affiliation{Department of Applied Science and Environment, 
University of Kochi, Kochi 780-8515, Japan}
\affiliation{Research Center for Nuclear Physics, Osaka University,
Ibaraki, Osaka 567-0047, Japan}
\date{\today}

\begin{abstract}
 We investigate the possibility of the existence of the exotic torus
 configuration in the high-spin excited states of $^{40}$Ca.  We here
 consider the spin alignments about the symmetry axis.  To this end, we
 use a three-dimensional cranked Skyrme Hartree-Fock method and search
 for stable single-particle configurations. We find one stable state
 with the torus configuration at the total angular momentum $J=$ 60
 $\hbar$ and an excitation energy of about 170 MeV in all calculations
 using various Skyrme interactions. The total angular momentum $J=60$
 $\hbar$ consists of aligned 12 nucleons with the orbital angular
 momenta $\Lambda=+4$, $+5$, and $+6$ for spin up-down neutrons and
 protons.  The obtained results strongly suggest that a macroscopic
 amount of circulating current breaking the time-reversal symmetry
 emerges in the high-spin excited state of $^{40}$Ca.
\end{abstract}

% insert suggested PACS numbers in braces on next line\
% insert suggested keywords - APS authors don't need to do this

%\maketitle must follow title, authors, abstract, \pacs, and \keywords
\maketitle

The investigations of nuclei rotating extremely fast about the symmetry
axis provide a good opportunity to severely test the fundamental theory
of a quantum mechanics.  In a classical picture for such rotation, the
nuclear density at the equatorial plane increases due to the strong
centrifugal force, and the oblate deformations develop with increasing
rotational frequencies~\cite{Cohen}.  However, such a collective
rotation about the symmetry axis is quantum-mechanically
forbidden. Instead, the spins and the orbital angular momenta of single
particles are aligned with the symmetry axis building extremely high
angular momenta~\cite{BM77,Afan99}.  In this case, the total angular
momentum of the nucleus is just the sum of the symmetry-axis components
of those of the single particles.  Bohr and Mottelson pointed out that
in such a nucleus, a ``macroscopic'' amount of circulating current
breaking the time-reversal symmetries emerges, which is a fascinating
new form of the nuclear matter~\cite{BM81}.

A typical example of such phenomena is the high-K oblate isomer states
in $^{152}$Dy~\cite{Khoo78}.  In the high angular-momentum states above
$I$ $=$ 14 $\hbar$, the observed excited states are irregularly
distributed around the average yrast line, which is calculated by
assuming a macroscopic nuclear shape with the strongly oblate
deformations ($\beta$ $\sim$ $-0.3$).  The emergence of these high-spin
isomers around the yrast line can be naturally explained by the
single-particle alignments~\cite{BM81}, because the E2 transitions,
which are a characteristic quantity related to the rotational
collectivity, are strongly forbidden for those nuclei. In addition, the
$\gamma$ transitions are strongly suppressed due to the large difference
of the internal structure between the initial and final states.  Those
high-spin isomers around the yrast line are thus called ``yrast traps''.
Many experiments have been attempted to produce yrast traps with
extremely high spins~\cite{Paul07}.  The observed highest angular
momentum in this category is $I$ = 49 $\hbar$ in
$^{158}$Er~\cite{Simp94}.

In this Letter, we investigate the possibility of the existence of the
{\it torus} configuration as an extreme limit of the
high-spin oblate isomers.  In such a limit, more nucleons with higher
orbital angular momentum are aligned and its density is much denser at
the equator part.  We here consider a special case for the existence of
the density only around its equator part. We below show that such exotic
state can indeed exist at an extremely high angular momentum in
$^{40}$Ca.

Many theoretical calculations have been performed to search for the high
spin states with strongly oblate deformations in a wide range of
nuclei. Those studies have been mainly focused on the nuclei with the
oblate deformation at around $\beta$ $\sim$ $-0.3$ heavier than
$^{40}$Ca~\cite{Afan99}.  In contrast, Wong discussed the stability of
the torus configuration using the macroscopic-microscopic model in
heavy-mass systems~\cite{Wong72,Wong73,Wong78}.  The calculations using
the constraint Hartree-Fock (-Bogoliubov) models have been also
performed for heavy-mass systems~\cite{Warda07,Sta09}.  The calculations
using $\alpha$ cluster model have been attempted to obtain the ring
configurations consisting of alpha particles and extra
neutrons~\cite{Wil86}.  However, as shown in the present study, an
essential mechanism for the stability of such configuration is the
extremely high angular momentum directed to the symmetry axis, resulting
in the breaking of the time-reversal symmetry in the intrinsic state.

To investigate the possibility of the existence of the torus
configuration in $^{40}$Ca with high angular momentum, we use the
three-dimensional Skyrme Hartree-Fock (HF) method with a Lagrange
multiplier, which is introduced to obtain the single-particle states
aligned with the symmetry axis.  To this end, we minimize the HF
Hamiltonian, $\hat{H}$, with the Lagrange multiplier, $\omega$.  In the
present study, we take the $z$ axis as the symmetry axis.  We first
define the effective Hamiltonian, $\hat{H'}$, given by
$\hat{H'}=\hat{H}-\omega \hat{J_z}$, where $\hat{J_z}$ denotes the
operator for the sum of the $z$ components of the total angular-momentum
for each single particle, $\hat{j_z}$, given by
$\hat{J_z}=\sum_{i}\hat{j_z^{(i)}}$.  In the HF approximation,
$\hat{H'}$ is rewritten as
$\hat{H'}=\sum_{i}\{\hat{h_i}-\omega\hat{j_z^{(i)}}\}$, where
$\hat{h_i}$ denotes the Hamiltonian for each single particle.  The
eigenvalue of $H'$ is given by
$\left<\hat{H'}\right>=\sum_{i}\{(e_i-\lambda)-\hbar\omega\Omega_i\}$,
where $\lambda$ denotes the Fermi energy at $\omega=$ 0 and $e_i$ and
$\Omega_i$ denote the energy and the $z$ component of the total angular
momentum in the unit of $\hbar$ for each single particle, respectively.
In the present study, we search for the stable state using the
equivalent cranked Skyrme HF equation, $\delta\left<\hat{H}-\omega
\hat{J_z}\right>=0$~\cite{Mar06,ichi11}, by scanning a large range of
$\omega$.

Before the HF calculations, we here discuss the shell structure of the
torus configuration using the radial displaced harmonic oscillator (RDHO)
model~\cite{Wong73}.  For the torus configuration, not only $\Omega$
but also the $z$ component of the orbital angular momentum, $\Lambda$, are
good quantum numbers ($\Omega=\Lambda+\Sigma$, where $\Sigma$ denotes
the $z$ component of the spin values, $\pm$1/2).
Two nucleons in each $\Lambda$ energetically degenerate with the
different spin values.  At $\hbar\omega=0$, the lowest configuration for
$^{40}$Ca is $\Lambda$ = 0, $\pm1$, $\pm2$, $\pm3$, and $\pm4$ and the
residual two nucleons can occupy any two states with $\Lambda=\pm5$.  At
$\hbar\omega\ne 0$, the possible spin aligned configurations are (i)
$\Lambda$ = 0, $\pm1$, $\pm2$, $\pm3$, $\pm4$ and $+5$ for the total angular
momentum $J=$ 20 $\hbar$ [= 5 $\hbar$ $\times$ 2 (spin degeneracy)
$\times$ 2 (isospin degeneracy)], (ii) $\Lambda$ = 0, $\pm1$, $\pm2$,
$\pm3$, $+4$, $+5$ and $+6$ for $J$ = 60 $\hbar$ [= 15 $\hbar$ $\times$ 2
$\times$ 2], and (iii) $\Lambda$ = 0, $\pm1$, $\pm2$, $+3$, $+4$, $+5$,
$+6$ and $+7$ for $J=$ 100 $\hbar$ [= 25 $\hbar$ $\times$ 2 $\times$ 2].

\begin{figure}[t]
\includegraphics[keepaspectratio,width=\linewidth]{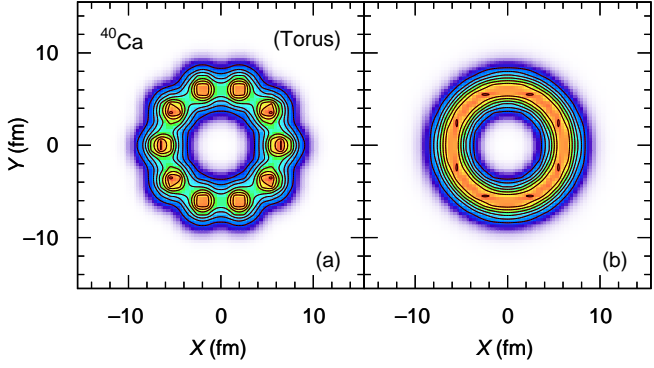}\\%
\caption{(color online) Total density for (a) the initial condition of the HF
 iterations and (b) the calculated result with $\hbar\omega=1.5$ MeV at the
 15000 HF iterations. The density is integrated in the $z$ direction. The
 contours correspond to multiple steps of 0.05 fm$^{-2}$. The color is
 normalized by the largest density in each plot.}
\end{figure}

 In the self-consistent calculations, the single-particle wave functions
are described on a Cartesian grid with a grid spacing of $1.0$ fm. We
take 32 $\times$ 32 $\times$ 24 grid points for the $x$, $y$, and $z$
directions, respectively.  This was sufficiently accurate to provide
converged configurations. The damped-gradient iteration method
\cite{gradient} is used, and all derivatives are calculated using the
Fourier transform method.  We take three different Skyrme forces which
all perform well concerning nuclear bulk properties but differ in
details: SLy6 as a recent fit which includes information on isotopic
trends and neutron matter \cite{Cha97a}, and SkI3 and SkI4 as recent
fits especially for the relativistic isovector structure of the
spin-orbit force \cite{Rei95a}.  However, except for the effective mass,
the bulk parameters (equilibrium energy and density, incompressibility,
and symmetry energy) are comparable in the all interactions.

For the initial wave functions, we chose the ring configuration with 10
$\alpha$ particles placed on the $x$-$y$ plane, as shown in Fig.~1(a).
Each $\alpha$ particle is described by the Gaussian function with its
center placed on $z=0$.  Using this initial condition, we perform the HF
iterations with 15000 times and investigate the convergence of the
calculated results.  Figure 2 shows the convergence behaviors of
$\left<\hat{J_z}\right>$ versus the number of the HF iterations with
various $\omega$'s using the SLy6 interaction. We can see that the
result calculated with $\hbar\omega=1.5$ MeV converges rapidly to
$J_z=60$ $\hbar$.  Figure 1(b) shows the density obtained with
$\hbar\omega=1.5$ MeV at the 15000th iteration step. The calculated
result is indeed the torus configuration. The obtained density
distribution, $\rho(r,z)$, can be well fitted by $\rho(r,z)=\rho_0
e^{-\{(r-r_0)^2+z^2\}/\sigma^2}$, where $\rho_0=0.13$ fm$^{-3}$,
$r_0=6.07$ fm, and $\sigma=1.61$ fm for the SLy6 interaction.
\begin{figure}[t]
\includegraphics[keepaspectratio,width=\linewidth]{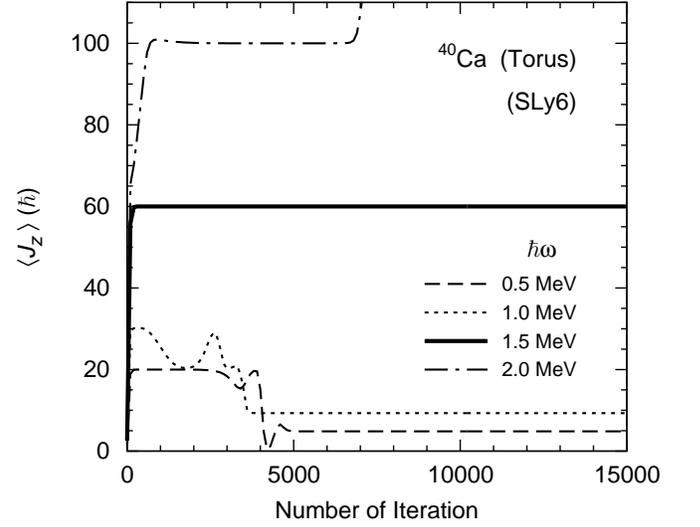}\\%
\caption{Convergence behavior of the expectation value of $J_z$ in the
 HF calculations versus the number of iterations.
 The dashed, dash-dotted, solid, and dotted lines denote the
 calculated results with $\hbar\omega=$ 0.5, 1.0, 1.5, and 2.0 MeV,
 respectively.} 
\end{figure}

On the other hand, the calculated results with $\hbar\omega=$ 0.5 and
1.0 MeV lead to unstable states. That for $\hbar\omega=2.0$ MeV leads to
the fission.  Although it seems that $\left<\hat{J_z}\right>$'s with
$\hbar\omega=$ 0.5 and 1.0 MeV converge at the 15000th step, those are
in fact unstable.  In Fig.~2, we can see that these states first
converge to a quasi-stable state. After that, the instability of those
states increases.  In those quasi-stable states, $J_z$ is 20 $\hbar$ for
$\hbar\omega=$ 0.5, and 1.0 MeV and is 100 $\hbar$ for $\hbar\omega=2.0$
MeV.  Later, we will discuss those quasi-stable states with $J_z=20$ and
100 $\hbar$.

\begin{figure}[t]
\includegraphics[keepaspectratio,width=\linewidth]{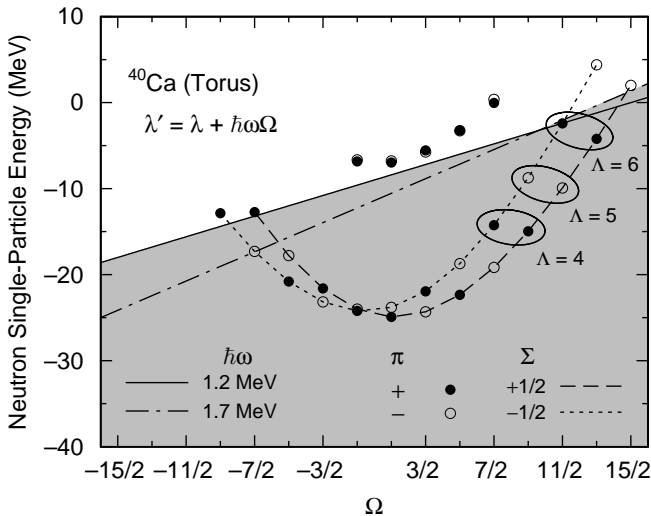}\\%
\caption{Single-particle energies versus
 the $z$ component of the total angular momentum $\Omega$. The solid and
 open circle denote the single-particle states with the positive and
 the negative parties, respectively. The single-particle states with the
 $z$ component of the spin $\Sigma=+1/2$ and $-1/2$ are connected by the
 dashed and dotted lines, 
 respectively. The solid and the dash-dotted lines denote the sloping 
 Fermi energies with $\hbar\omega=1.2$ and 1.7 MeV, respectively. The
 gray area denotes the states below the Fermi energy.} 
\end{figure}

We next investigate the region of the Lagrange multipliers where the
torus configurations stabilize, using various Skyrme interactions.  We
find that such region of $\hbar\omega$ for the SLy6 and SkI3
interactions extends from 1.2 to 1.7 MeV. That for the SkI4 interaction
is from 1.2 to 1.8 MeV.  In such regions, we obtain the only one stable
torus configuration at $J_z=60$ $\hbar$ in all calculations. The
obtained excitation energies are 170.45, 174.22, and 172.53 MeV for the
SLy6, SkI3, and SkI4 interactions, respectively.  The fitted radius of
the torus configuration is almost the same for all calculations. The
dependence of the calculated result on the choice of the interaction is
weak.  This is mainly because that the spin-orbit forces in the radial
directions of the inner and outer surface of the torus configuration
almost cancel out each other (see Eq.~(4.14) in
Ref.~\cite{Wong73}). Thus, the RDHO mode is a good approximation for the
present calculations.

Figure 3 shows the single-particle spectrum versus $\Omega$ calculated
using the SLy6 interaction. The solid and open circles denote the
single-particle states with the positive and the negative parities,
respectively.  The single-particle states with $\Sigma=$ $+1/2$ and
$-1/2$ are connected by the dashed and the dotted lines,
respectively. These energies connected by each line are well
proportional to $\Lambda^2$, which is consistent with that of the RDHO
model [see Eq.~(4.26) in Ref.~\cite{Wong73}].  We here define the
``sloping'' Fermi energy, $\lambda'$, given by
$\lambda'=\lambda+\hbar\omega\Omega$~\cite{Afan99}. Then, the occupied
states are obtained by $\left<\hat{H'}\right>=\sum_{i}(e_i-\lambda')$
with $e_i<\lambda'$.  The solid and the dash-dotted lines denote the
sloping Fermi energies with $\hbar\omega=1.7$ and 1.2 MeV,
respectively. The occupied states are denoted by the gray area. In
Fig.~3, we can clearly see that the torus configuration can exist at
$J_z=60$ $\hbar$ in the region from $\hbar\omega=1.2$ to 1.7 MeV.  The
total angular momentum of $J_z=60$ $\hbar$ consists of the spin-up-down
pairs of the aligned single particles with the same $\Lambda$,
($\Omega_+$, $\Omega_-$), of ($+9/2^+$, $+7/2^+$) with $\Lambda=4$,
($+11/2^-$, $+9/2^-$) with $\Lambda=5$, and ($+13/2^+$, $+11/2^+$) with
$\Lambda=6$ (see the solid ellipses in Fig.~3).

\begin{figure}[t]
\includegraphics[keepaspectratio,width=\linewidth]{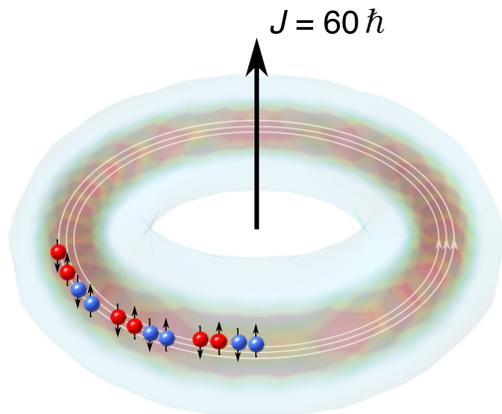}\\%
\caption{(color online) Circulating currents of 12
nucleons. The outer layer of the torus surface denotes the half of the
total density. The density plot in the inside of the torus surface
denotes the calculated density of the aligned 12 single particles.}
\end{figure}

As discussed in the shell structure of the torus configuration, the
states with $J=20$, 60, and 100 $\hbar$ are the possible combinations of
the spin alignment for the RDHO model.  For the quasi-stable state with
$J=20$ $\hbar$, the centrifugal force is insufficient to stabilize the
torus configuration against the strong nuclear attractive force. On the
other hand, to build the state with $J=100$ $\hbar$, the unbound states
($e>0$) with $\Omega=+15/2$ and $+11/2$ are occupied, resulting in the
instability for the torus configuration. Only for $J=60$ $\hbar$, the
torus configuration can be stabilized.  Thus, the stability of the state
with $J=60$ $\hbar$ is rather robust.  Although we have performed the
similar calculations for $^{24}$Mg and $^{32}$S, we could not find the
stable torus states for those nuclei.

As shown in the present study, if the state with the torus configuration
is formed at $J=60$ $\hbar$, the macroscopic circulating current of 12
nucleons with $\Lambda=+4$, +5, and +6 strongly violates the
time-reversal symmetry in the intrinsic state (see Fig.~4).  It is
interesting to investigate how this fascinating new state can be
observed in experiments.  The state with the torus configuration at
$J=60$ $\hbar$ would have an extremely large magnetic moment
($\mu=30\mu_N$). This would lead to a procession motion under an
external magnetic field.

A question then arises how such a ``femto-scale magnet'' rotates
spontaneously.  When the spherical symmetry is broken, the collective
rotation emerges spontaneously, in principle, to restore the broken
symmetry.  However, it is unclear whether such a state built with
significant amount of circulating current can rotate about the
perpendicular direction to the symmetry axis or not.  If not, such a
state would be an anomalous one, which has not yet been recognized in
the experiments.  Even if the state can rotate, the rotational band
built on such a state would show interesting behaviors for their M1/E2
transition strengths and moment of inertia, $\mathscr{T}_\perp$.  It is
intriguing to investigate the extent of the difference between the
experimental and the classical rigid-body moments of inertia
($\mathscr{T}_\perp=$ 20.97 $\hbar^2\cdot$MeV$^{-1}$ in the case of the
SLy6 interaction).

To identify the torus configuration in the experiment, it is also
important to discuss the competition to the fission decay channel.  The
calculation using the macroscopic model with the FRLDM2002 parameter
set~\cite{FRLDM2002} shows that the fission barrier for the spheroidal
deformations vanishes at $J=32$ $\hbar$ and an excitation energy of
49.27 MeV~\cite{Sierk86}. Our calculated results for the torus
configuration are considerably higher than those values. However, it
would be possible that the state with the torus configuration can
survive against the fission.  One important example for such possibility
is the long-lived $K^{\pi}=16^+$ isomer in $^{178}$Hf with a half-lives
of about 31 y~\cite{Khoo76}, which is extremely long compared to other
isomers (4 s for the $K^{\pi}=8^-$ isomer).  In this connection, it is
unclear how the collective path from a topologically different torus
configuration is connected to the region of the spheroidal deformations
leading to the fission~\cite{Wong78,Sta09}.  Wong indeed showed that the
torus configurations are stable from $J\sim57$ to 74 $\hbar$ for mass
number $A\sim50$~\cite{Wong78}.

In summary, we have suggested the existence of the torus configuration
 in the extremely high-spin excited states of $^{40}$Ca using the
 three-dimensional cranked Skyrme HF method.  We found only one stable
 state with the torus configuration at $J_z=60$ $\hbar$ in all the
 calculations with any Skyrme interactions. The calculated excitation
 energies of this state are 170.45, 174.22, and 172.53 MeV for the SLy6,
 SkI3, and SkI4 interactions, respectively.  The obtained results are
 insensitive to the choice of the Skyrme interactions, because the
 contribution of the spin-orbit force is small in the torus
 configuration.  To build the torus state with $J=60$ $\hbar$, the 6
 nucleons for $\Omega=$ $+9/2^+$ and $+7/2^+$ with $\Lambda=4$,
 $\Omega=$ $+11/2^-$ and $+9/2^-$ with $\Lambda=5$, and $\Omega=$
 $+13/2^+$ and $+11/2^+$ with $\Lambda=6$ are aligned with the symmetry
 axis for both protons and neutrons.  We have shown that such
 configuration is built on the major shell structure estimated by the
 RDHO model for the rotating torus shape, indicating that the torus
 state is robustly stable and the macroscopic amount of the
 time-reversal symmetry breaking occurs.  Although the observation of
 the torus state would be difficult, exploration for such an exotic
 state would provide us valuable information on the new frontier of the
 nuclear matter, which is a big challenge both theoretically and
 experimentally.

\begin{acknowledgments}
 A part of this research has been funded by MEXT HPCI STRATEGIC PROGRAM.
  This work was undertaken as part by the Yukawa International Project
  for Quark-Hadron Sciences (YIPQS), and was partly supported by the
  GCOE program ``The Next Generation of Physics, Spun from Universality
  and Emergence'' from MEXT of Japan.  J.A. M. and P.-G. R. were
  supported by BMBF under contract numbers 06ER9063 und 06FY9086,
  respectively.  One of the authors (JAM) would like to thank the Japan
  Society for the Promotion of Science (JSPS) for an invitation
  fellowship for research in Japan.
\end{acknowledgments}

\end{document}